\documentclass[aps,pra,preprint,groupedaddress,twocolumn,10pt]{revtex4-1}
\usepackage{graphicx}
\usepackage{epsfig}
\usepackage{epstopdf}
\usepackage{subfigure}
\usepackage{appendix}
\begin{document}
\title{Estimation of temporal separation of slow light pulses in atomic vapors by weak measurement}
\author{Pardeep Kumar}

\email[]{pradeep.kumar@iitrpr.ac.in}
\author{Shubhrangshu Dasgupta}

\affiliation{Department of Physics, Indian Institute of Technology Ropar, Rupnagar, Punjab 140001, India}

\date{\today}

\begin{abstract}
We show how two circular polarization components of a linearly polarized pulse, propagating through a coherently driven dilute atomic vapor, can be well resolved in time domain by weak measurement. Slower group velocity of one of the components due to electromagnetically induced transparency leads to a differential group delay between the two components. For low number density, this delay may not be large enough to temporally resolve the two components. We show how this can be enhanced in terms of mean time of arrival of the output pulse through a post-selected polarizer. We demonstrate the idea with all the analytical and numerical results, with a specific example of alkali atoms.
\end{abstract}


\maketitle

\section{Introduction}
It is known from the measurement theory that measurement of an observable of the system over a long period of time leads to a statistical average of all the output results put together (called ``the expectation value'' of the observable). This is valid irrespective of the interaction strength between the system and the measuring device. However, in case of weak interaction, contrary to strong interaction, the system does not collapse into one of the eigenstates of the observable, leading to a large uncertainty at the output. In such a case, a large ``weak value'' of the observable can be obtained, if one considers only a post-selected set of the output results. Such a post-selection of the basis for strong measurement immediately after weak interaction  makes an essential component of the weak measurement procedure which was  proposed in \cite{ahar1988} by Aharonov, Albert, and Vaidman and further investigated in \cite{ahar1989,duck1989,ahar1990}. The technique of weak measurement was first demonstrated in optical experiment by Ritchie \textit{et al.} \cite{hulet1991}. Weak value has been demonstrated to be useful in  small parameter estimation, e.g., in resolving Angstrom-scale optical beam deflection \cite{hoston2008,dixon2009}, frequency shifts \cite{starling2010}, phase shifts \cite{david2010}, temporal shifts \cite{brunner2010,bruder2013} and temperature shifts \cite{stone2012}. It has also been employed in the direct measurement of quantum states \cite{lundeen2011,lundeen2012,lundeen2014}. A nice review on weak measurement can be found in \cite{boyd2014}. Recently, the weak measurements are also proposed in quantum dots  \cite{williams2008, romito2008}.
 
Previous demonstrations of weak measurement in optical experiments \cite{hulet1991,hoston2008,starling2010,david2010,brunner2010,bruder2013} can be explained semiclassically using a wave equation derived from Maxwell's equations. Recently, the experiments have been proposed \cite{ahnert2004,agarwal2007,hill2008,simon2011} and demonstrated \cite{pryde2005,wang2006,howell2010} which can be explained with quantum mechanical perspective. In \cite{pryde2005}, an entangling circuit has been shown to  enable one single photon to make weak measurement of the polarization of the other. Further, weak values of the observables using entangled photons in parametric down-conversion have been explored in \cite{agarwal2007}. Method to generate two qubit entanglement in a controlled way, using weak measurements has been proposed in \cite{hill2008}.  The quantum mechanical explanation for the interferometric weak value deflections \cite{dixon2009} has been given in \cite{howell2010}.  Wang \textit{et al.} \cite{wang2006} have presented an experiment to measure the weak values of the arrival time of a single photon, by using the proposal of Ahnert \cite{ahnert2004}.

The usefulness  of weak values has also been perfectly demonstrated in the superluminal propagation of charged particle \cite{daniel2002} and of electromagnetic pulse  propagation \cite{solli2004,brunner2004}. 
Solli \textit{et al.} \cite{solli2004} in the experiment with polarized \textit{microwaves} and two-dimensional birefringent \textit{photonic crystals}, have derived a complex relation between system's response function and the weak values of the polarization of the photon. On the other hand, in an experiment with \textit{optical pulses} in \textit{optical fiber}, Brunner \textit{et al.} \cite{brunner2003,brunner2004} have shown that weak value of polarization can also be related to the mean time of arrival $(\langle t\rangle)$ of the pulse in a postselected polarization, as

\begin{equation}
\langle t\rangle = \frac{\delta \tau}{2}\mbox{Re}\langle \sigma_{z}\rangle_{w}
\label{Eq. 1}
\end{equation}
where, $\delta\tau$ is the differential group delay (DGD), i.e., the temporal separation between the pulse peaks of the polarization modes $(|H\rangle, |V\rangle)$, which are eigenstates of the operator $\sigma_{z}$: $\sigma_{z}|H\rangle = |H\rangle$ and  $\sigma_{z}|V\rangle = -|V\rangle$. Here,
\begin{equation}
 \mbox{Re}\langle \sigma_{z}\rangle_{w} = \mbox{Re}\left[\frac{\langle \psi_{f}|\sigma_{z}|\psi_{i}\rangle}{\langle \psi_{f}|\psi_{i} \rangle}\right]
 \label{Eq. 2}
 \end{equation}
 is the real part of the weak value of the polarization observable when it is measured between a preselected state $|\psi_{i}\rangle$ and a postselected state $|\psi_{f}\rangle$.

 In this paper, we employ the above concept of mean time of arrival and weak value (Eq. (\ref{Eq. 1})) to an \textit{optical pulse} with preselected polarization, propagating through the \textit{atomic vapor systems} \cite{itay2013,nori2013}. Such systems, with suitable atomic configuration and in presence of a strong coherent field, can act as a polarization splitter of pulses \cite{dasgupta2002}.  This happens when two polarization components propagate through the atomic medium with negligible absorption, but with different group velocities. We show, if the two circularly polarized components of a linearly polarized probe pulse are not well resolved in time domain, their temporal separation $\delta\tau$ can be indirectly inferred by measuring the mean time of arrival  $\langle t\rangle$ of a post-selected pulse. Note that $\langle t\rangle$ can be made much larger than $\delta\tau$ by a suitable choice of the post-selected basis. Therefore, the output pulse can be well resolved from the reference input pulse. We further obtain an one-to-one correspondence between $\delta\tau$ and $\langle t\rangle$ over large range of the control parameters.  Such kind of correspondence can be used to obtain the number density $(N)$ of the atomic medium, as well. Note that such idea was employed in \cite{brunner2004} to temporally resolve two photon pulses in optical fibers and indirectly calculating their superluminal velocities,  while in this paper, we demonstrate this using slow optical pulses propagating through a coherently driven atomic vapors. In addition, we can also coherently control the time-separation $\delta\tau$ and therefore the weak values, a feature that is inherently absent in optical fibers. It is to be noted here that a semiclassical treatment is sufficient to explain the results presented in this paper. 

The structure of the paper is as follows. In Sec. II, we theoretically study the temporal separation between the circular components of the probe pulse via weak measurements, which can be found through appropriate preselected and postselected states of the system. In this section the main results of the paper are presented. Sec. III contains the discussion and concluding remarks.

\section{Weak measurement in atomic vapors}
 Three necessary steps for weak measurements are: (1) preselection of a quantum state; (2) weak interaction between the system and the measuring device; (3) postselection of a quantum state. We consider weak measurement of polarization of an optical pulse and the energy of the pulse takes the role of the measuring device. We describe below, how this measurement can be done using three essential steps mentioned above.
\subsection{Preselection of Polarization}
We consider the propagation of a linearly polarized input probe pulse through an atomic medium (with two ground states $|1\rangle$ and $|2\rangle$ and two excited states $|3\rangle$ and $|4\rangle$; see Appendix for detailed discussion) of length $L$. The input pulse can be written in terms of its Fourier components as 
\begin{equation}
\vec{E}(z,t) = \hat{x}\int_{-\infty}^{+\infty}\varepsilon(\omega)\exp\left\{i\omega\left(\frac{z}{c}-t\right)\right\}d\omega + c.c.
\label{Eq. 3}
\end{equation}
where, $\varepsilon(\omega)$ is the amplitude of profile of the pulse. Here  $\hat{x}$ denotes the polarization state of the probe pulse and can be written as 
\begin{equation}
|\psi_{o}\rangle = \frac{1}{\sqrt{2}}(|\sigma_{+}\rangle + |\sigma_{-}\rangle),
\label{Eq. 4}
\end{equation}
which is equivalent to the preselection of polarization. Note that the circular polarization states $(|\sigma_{+}\rangle,|\sigma_{-}\rangle)$ make the eigen-basis of the Pauli spin-matrix $\sigma_{y}$:  $\sigma_{y}|\sigma_{\pm}\rangle = \pm|\sigma_{\pm}\rangle$. 

In presence of the control field, these two polarization components propagate with different group velocities through the atomic medium. This can be explained as follows. When the central frequency of the probe pulse is in near resonance with $|4\rangle \leftrightarrow |1\rangle$ transition, the control field, resonant with the $|4\rangle\leftrightarrow |2\rangle$ transition makes an electromagnetically induced transparency (EIT) window for the $\sigma_-$ polarization component. This is also associated with a large normal dispersion and hence a slow group velocity of this component. A magnetic field can be applied to remove the degeneracy of the ground and the excited state manifold. This makes the $\sigma_{+}$ component far from resonance from the $|3\rangle\leftrightarrow |2\rangle$ transition. Therefore the dispersion profile for this component exhibits a much flatter behavior in the frequency domain, leading to its group velocity not too different from $c$, the velocity of light in vacuum. Because of the difference in the group velocities inside the medium, the two circularly polarized components come out of the medium at different times with negligible absorption. In this way, the medium temporally separates the two orthogonal polarization components of the input pulse.
\subsection{Weak Interaction}
 In this paper, we focus on a situation where the temporal separation $\delta\tau$ between circular polarized components is much smaller than the width of the pulse such that the detector cannot resolve between these two components in time-domain i.e. $(\delta\tau < \Delta t)$, $\Delta t$ being the temporal resolution of the detector. Because, the group velocities are proportional to the number density $N$ of the atomic medium, such a situation would arise when $N$ is sufficiently small. In this paper,  we consider $N = 10^{9} \mbox{cm}^{-3}$.  Note that this situation mimics a weak interaction between the light pulse and the atomic medium, that plays the role of a measuring device for the polarization state of the pulse. 
 
For negligible absorption, the weak interaction gives rise to a rotation of polarization of the input field of frequency $(\omega_{0})$ and the polarization state of the output field would read as \cite{brunner2004} 
\begin{equation}
|\phi\rangle =  \frac{1}{\sqrt{2}}\left[e^{i\delta\tau \omega_{o}/2}|\sigma_{+}\rangle + e^{-i\delta\tau \omega_{o}/2}|\sigma_{-}\rangle\right]
\label{Eq. 5}
\end{equation}

In the present case, to demonstrate the weak interaction, we choose the following Gaussian profile for the input pulse Eq. (\ref{Eq. 3}), as shown in Fig. 1(a):
\begin{equation}
\varepsilon(\omega) = \varepsilon_{0}\frac{1}{\sigma \sqrt{\pi}}\mbox{exp}\left[-\omega^{2}/\sigma^{2}\right];
\varepsilon(t) = \varepsilon_{0}\frac{1}{\sqrt{2\pi}}\mbox{exp}\left[-\sigma_{t}^{2}t^{2}\right]
\label{Eq. 6}
\end{equation}
where, $\sigma$ ($\sigma_{t} = 2/\sigma$) is the width of the Gaussian pulse in frequency (time) domain and $\varepsilon_{0}$ is the pulse amplitude. 
\begin{figure}[!ht]
\begin{center}
\begin{tabular}{ll}
\subfigure[]{\includegraphics[scale=0.12]{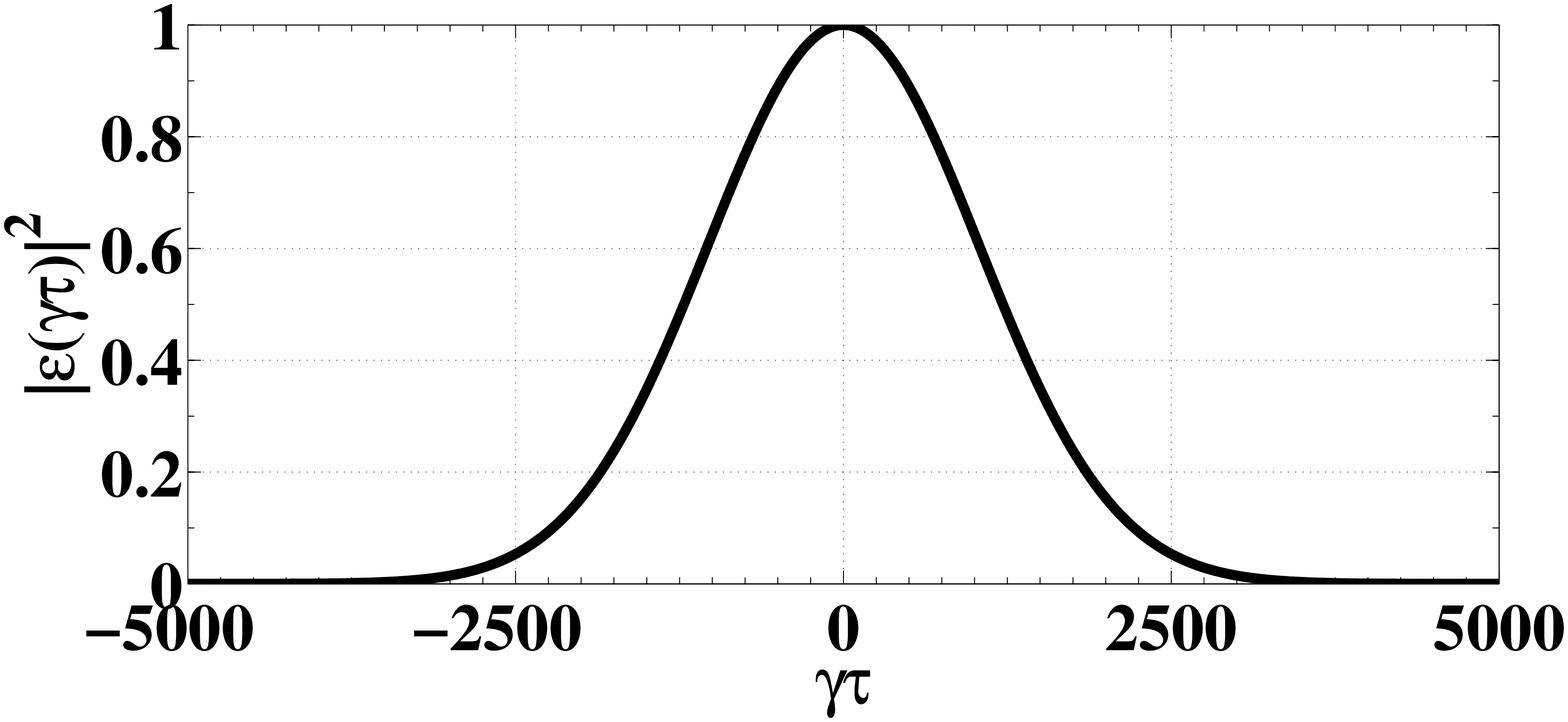}} & \subfigure[]{\includegraphics[scale=0.12]{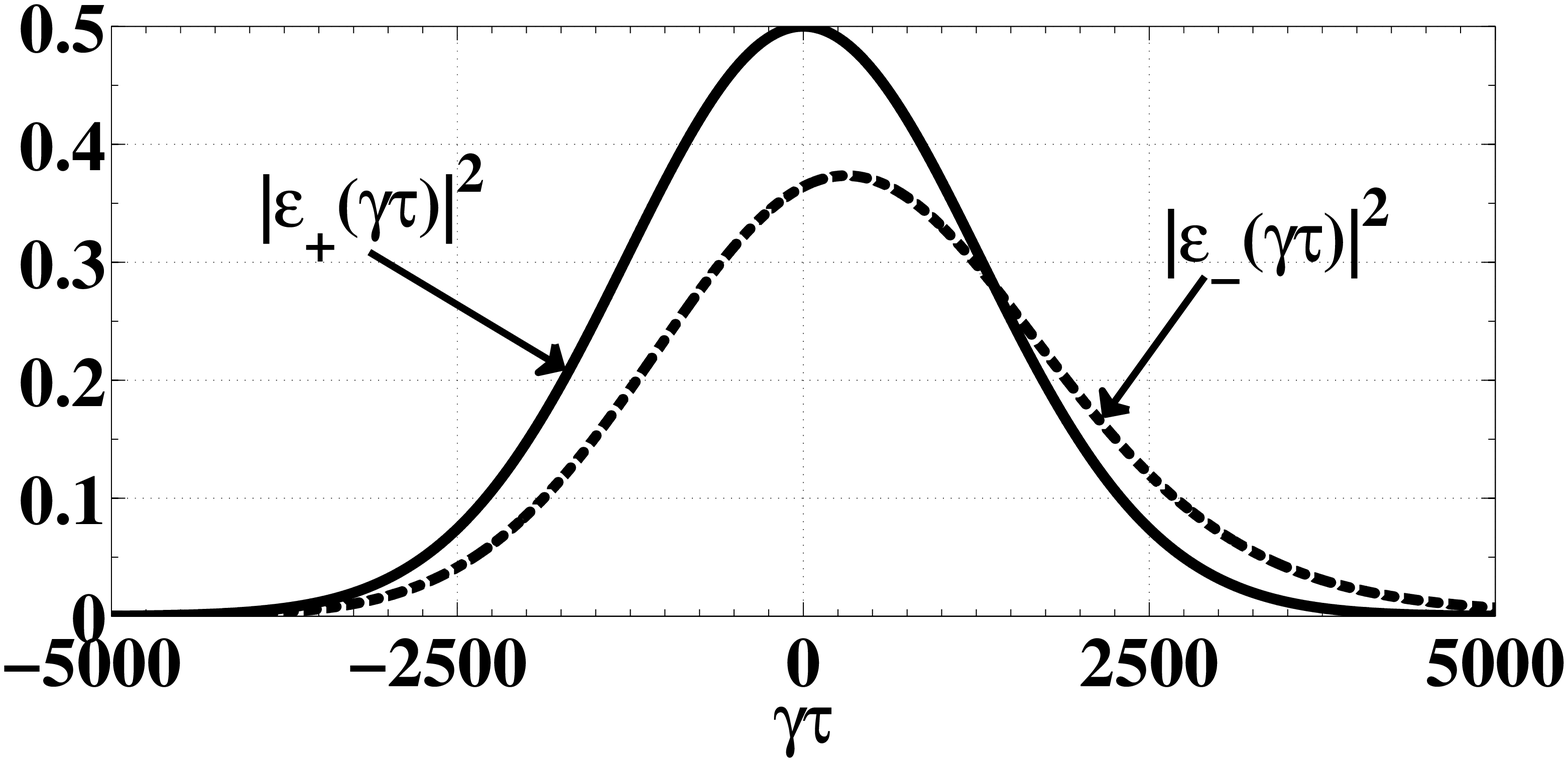}}
\end{tabular}
\begin{center}
\caption{(a) The input Gaussian pulse in time domain with a width of $\sigma_{t} = $400$~\mu$s ($\sigma$ = 2$\pi$ $\times$ 0.7916 kHz) (b) The circular polarization components [$\sigma_{+}$ (solid line) and $\sigma_{-}$ (dotted line)] at the output of the medium. We choose the following parameters: $N = 10^{9}$ atoms cm$^{-3}$, the length of the medium $L = 1$ cm, central wavelength of the input pulse $\lambda = 769.9$ nm, the spontaneous emission rate $A=2\pi \times 6.079$ MHz, pertaining to $^{39}$K atoms, $\gamma = \frac{A}{6}$,  $B = 5\gamma$, $\gamma_{13} = \gamma_{24} = \gamma$, $\gamma_{23} = \gamma_{14} = 2\gamma$, $\gamma_{coll} = 0$, $\Gamma_{32} = \Gamma_{31} = \Gamma_{41} = \Gamma_{42} = \frac{3}{2}\gamma, \Gamma_{43} = 3\gamma, \Gamma_{21} = \gamma_{coll}$, $G = 0.03\gamma$  and $\tau = t - L/c$.}
\end{center}
\label{fig1}
\end{center}
\end{figure}

The weak interaction of the linearly polarized probe pulse induce the following polarization in the medium
\begin{equation}
\vec{P}(z,t) = P_{+}(z,t)|\sigma_{+}\rangle +  P_{-}(z,t)|\sigma_{-}\rangle
\label{Eq. 7}
\end{equation}
where,
\begin{equation}
P_{\pm}(z,t) = \int_{-\infty}^{+\infty}\chi_{\pm}(\omega)\varepsilon(z,\omega)e^{-i\omega t}d\omega
\label{Eq.8}
\end{equation}
Here $\chi_\pm(\omega)$ are the complex susceptibilities of the atomic medium corresponding to the $\sigma_\pm$ components and can be calculated using Eqs. (A.10) and (A.11) in the Appendix. Further $\varepsilon_{\pm}(\omega) = \frac{\varepsilon(\omega)}{\sqrt{2}}$ are the amplitudes for $\sigma_{\pm}$ components and $\varepsilon(\omega)$ is given by Eq. (\ref{Eq. 6}).

 The polarization state of the output field (at $z = L$) after the weak interaction reads as
\begin{equation}
|\psi\rangle = E_{+}(t)|\sigma_{+}\rangle + E_{-}(t)|\sigma_{-}\rangle
\label{Eq. 9}
\end{equation}
where,

\begin{equation}
E_{\pm}(t) = \int_{-\infty}^{+\infty}d\omega \varepsilon_{\pm}(\omega)\exp\left\{ i\omega \left(\frac{L}{c} - t\right) + \frac{2\pi i\omega L}{c}\chi_{\pm}(\omega)\right\}\;.
\label{Eq. 10}
\end{equation}

Note that the pulse  suffers finite absorption (though small due to EIT condition for $\sigma_-$ component and off-resonance of the other), contrary to the case of optical fiber, which would gives rise to the polarization as described in Eq. (\ref{Eq. 5}).

We display in Fig. 1(b) the output intensities of both the polarization components. It is evident that these two components are not well resolved in time-domain. We find the temporal separation to be $\delta\tau = 305/\gamma$, while the pulse width in the time-domain is $\sigma_{t} = 2520/\gamma$ $(\gg \delta\tau)$. It is clear that, one of the polarization components suffers absorption.
\subsection{Post-selection of Polarization}

\begin{figure}[h!]
\begin{center}
\includegraphics[scale=0.20]{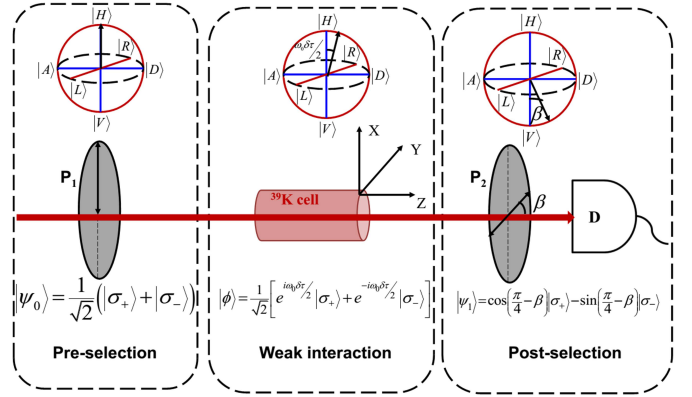}
\caption{Schematic diagram of the setup. Polarizer $P_{1}$ performs the pre-selection of polarization state, polarizer $P_{2}$ performs the post-selection and $D$ is the detector. Poincar\'{e} spheres represent the polarization states of the fields at each stage.}
\label{fig2}
\end{center}
\end{figure}

 The final and most important step of the weak measurement, is a post-selection of measurement basis. This is equivalent to using a polarizer with suitable orientation at the output. We choose the following post-selected polarization basis:

\begin{equation}
|\psi_{1}\rangle = \cos \left(\frac{\pi}{4}-\beta \right)|\sigma_{+}\rangle - \sin \left(\frac{\pi}{4}-\beta \right)|\sigma_{-}\rangle
\label{Eq. 11}
\end{equation}
where $0 < \beta \ll 1$. This makes $|\psi_1\rangle$ nearly orthogonal to $|\psi_o\rangle$. The schematic optical set-up of all the above three essential steps of weak measurement is shown in Fig. \ref{fig2}. 

The temporal shape of the output pulse through this polarizer can thus be written as 
\begin{equation}
\begin{array}{ccc}
\vec{E}_{out}(t) &=& \left[\cos \left(\frac{\pi}{4}-\beta \right)\langle\sigma_{+}| - \sin \left(\frac{\pi}{4}-\beta \right)\langle\sigma_{-}| \right]\\
& \cdot&\hspace{-2.45cm} \left[E_{+}(t)|\sigma_{+}\rangle + E_{-}(t)|\sigma_{-}\rangle \right]
\label{Eq. 12}
\end{array}
\end{equation}
Therefore the output intensity $I_{out}(t)$ = $|\vec{E}_{out}(t)|^{2}$ reads as
\begin{equation}
\begin{array}{ccc}
I_{out}(t) &=& \left|E_{+}(t)\right|^{2}\cos^{2}\left(\frac{\pi}{4}-\beta\right)+\left|E_{-}(t)\right|^{2}\sin^{2}\left(\frac{\pi}{4}-\beta\right)\\
&&-2\mbox{Re}\left(E^{\ast}_{+}(t)E_{-}(t)\right)\sin \left(\frac{\pi}{4}-\beta \right)\cos \left(\frac{\pi}{4}-\beta \right)
\end{array}
\label{Eq. 13}
\end{equation}
where, $\left|E_{\pm}(t)\right|^{2}$  correspond to the Gaussian profiles for $|\sigma_{\pm}\rangle$ components. Note that the strong interaction refers to the fact that two polarization components are well resolved, i.e., $\delta\tau \gg \sigma$. This is because, the interference term in Eq. (\ref{Eq. 13}) almost vanishes at this limit, i.e. $E^{\ast}_{+}(t)E_{-}(t)\approx 0$. But for the case of weak interaction, $\delta\tau \ll \sigma$, the two Gaussian pulses overlap as shown in Fig. 1(b) and the contribution of the interference term to the output intensity is no longer negligible. This leads to a very different temporal profile of the output pulse, as discussed below.  
\subsection{Temporal separation in terms of Weak values}
To obtain temporal separation for these two overlapping pulses at the output, we make use of the weak value of the polarization, defined as \cite{brunner2003}
 \begin{equation}
W = \frac{\langle \psi_{1}|\sigma_{y}|\psi \rangle}{\langle \psi_{1}|\psi\rangle}
\label{Eq. 14}
\end{equation}
 As seen from Eq. (\ref{Eq. 14}), $W$ can be made arbitrarily large by choosing the state $|\psi\rangle$ and  $|\psi_{1}\rangle$ nearly orthogonal, i.e., $\langle \psi_{1}|\psi\rangle \approx 0$. This weak value is related with mean time of arrival through Eq. (\ref{Eq. 1}), that can be calculated through the following relation:
\begin{equation}
\langle t\rangle = \frac{\int tI_{out}dt}{\int I_{out}dt}
\label{Eq. 15}
\end{equation}
Using (\ref{Eq. 9}), (\ref{Eq. 11}), and (\ref{Eq. 14}), we have
\begin{equation}
\mbox{Re}W  = \frac{\cot(\beta)}{\cos^{2}\left(\frac{\omega_{0}\delta\tau}{2}\right)+\cot^{2}(\beta)\sin^{2}\left(\frac{\omega_{0}\delta\tau}{2}\right)}
\label{Eq. 16}
\end{equation}
From (\ref{Eq. 1}), we finally obtain the following relation between $\delta\tau$ and $\langle t\rangle$:
\begin{equation}
\langle t \rangle = \frac{\delta \tau}{2}\mbox{Re}\langle \sigma_{y}\rangle_{w} = \frac{\delta\tau}{2}\cdot \frac{\cot(\beta)}{\cos^{2}\left(\frac{\omega_{0}\delta\tau}{2}\right)+\cot^{2}(\beta)\sin^{2}\left(\frac{\omega_{0}\delta\tau}{2}\right)}
\label{Eq. 17}
\end{equation}

With increase in the Rabi frequency $2G$ of the control field, the slope of the dispersion profile decreases, leading to a decrease (towards $c$) in the group velocity  of the $\sigma_-$ polarization component. Therefore, the temporal separation $\delta\tau$ between $\sigma_{+}$ and $\sigma_{-}$ components decreases. In this regime, the denominator of Eq. (\ref{Eq. 17}) approaches to unity and the mean time of arrival becomes $\langle t\rangle \approx \frac{\delta\tau}{2}\cot(\beta)$. 

\begin{figure}[!ht]
\includegraphics[scale=0.23]{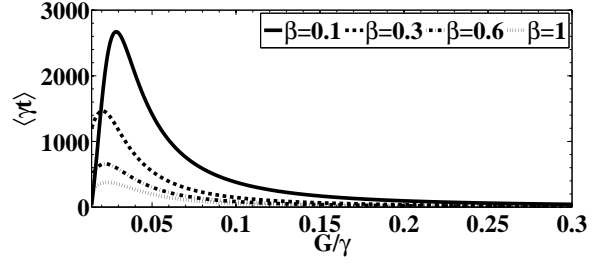}
\caption{Variation of mean time of arrival with $G/\gamma$ for different values of $\beta$. Other parameters are same as in Fig. 1. }
\label{fig3}
\end{figure}

As we choose $\beta\ll 1$, the above discussion suggests that $\langle t\rangle$ can be made much larger than $\delta\tau$. Thus, the temporal separation can be inferred indirectly by measuring the mean time of arrival of the output pulse through a post-selected polarizer. This is possible by carefully choosing the relative orientation of the pre-selected and the post-selected polarizer, i.e., by suitably choosing $\beta$. It can be observed from Fig. \ref{fig3} that at $\beta = 0.1$, the mean time of arrival attains the highest value. We choose $\beta = 0.1$ in the rest of the paper.
\begin{figure}[!ht]
\begin{center}
\includegraphics[scale=0.23]{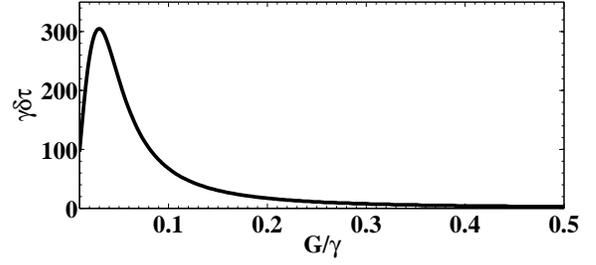}
\caption{The variation of differential group delay $(\gamma\delta\tau)$ with $G/\gamma$ . The other parameters are same as in Fig. \ref{fig3}.}
\label{fig4}
\end{center}
\end{figure}
\begin{figure}[!ht]
\includegraphics[scale=0.23]{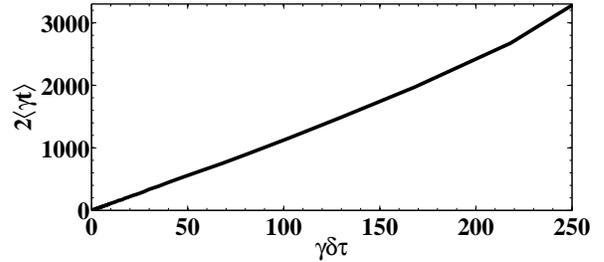}
\caption{The variation of 2$\langle \gamma t\rangle$ with the $\gamma\delta\tau$. The parameters used here are same as in Fig. \ref{fig4}.}
\label{fig5}
\end{figure}

Further, an one-to-one correspondence between $\delta\tau$ and $\langle t\rangle$ is evident by comparing the Figs. \ref{fig3} and \ref{fig4}. For a given value of  $G/\gamma$, the value of $\langle \gamma t\rangle$ and the corresponding value of the temporal separation  $\gamma\delta\tau$ can be inferred from Figs. \ref{fig3} and \ref{fig4}.   We demonstrate this correspondence in Fig. \ref{fig5} where we show how 2$\langle \gamma t\rangle$ varies linearly with $\gamma\delta\tau$. In this linear regime, the absorption of the two polarization components is negligible and the slope of this linearity is approximately equal to $\frac{\cot\beta}{2}$. Note that in non-linear regime, $\langle \gamma t\rangle$ depends non-trivially on $G/\gamma$ (in addition to $\beta$, see Eq. (\ref{Eq. 17})) and thus control field provides an extra handle in measuring $\langle \gamma t\rangle$.  \\
\begin{figure}[!ht]
\includegraphics[scale=0.23]{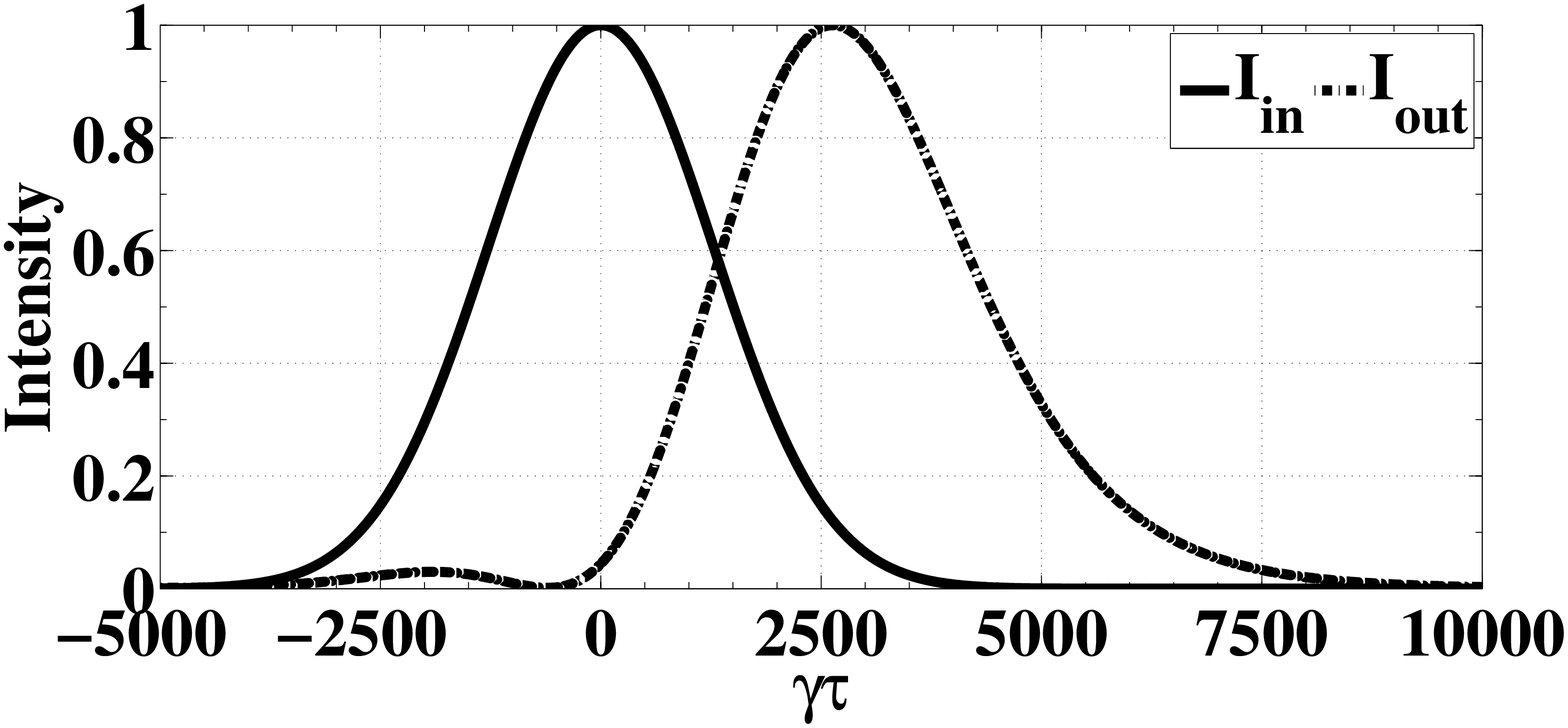}
\caption{The input Gaussian pulse (solid line) and the normalized output pulse (dot-dashed line). Here, $\beta = 0.1$ and rest of the parameters used are same as in Fig. 1. The output intensity of the probe pulse corresponds to 9.32 nW/cm$^{2}$ for input intensity of 1 $\mu$W/cm$^{2}$.}
\label{fig6}
\end{figure}
We demonstrate in Fig. \ref{fig6}, how the use of a post-selected polarizer with $\beta=0.1$ leads to a large temporal shift of the output pulse, compared to the input reference pulse (the peak-to-peak separation between these two pulses is found to be 2618/$\gamma$). We also calculate from Eq. (\ref{Eq. 15}) the mean time of arrival to be 2610/$\gamma$, that is in good agreement with Fig. \ref{fig6}. Note that here $\langle \gamma t\rangle > \sigma_{t}$ refers to a very good improvement of the resolution between two circular components [$\gamma\delta\tau \ll \sigma_{t}$, as in Fig. 1(b)].  The value of temporal separation obtained from Eq. (\ref{Eq. 17}) is then $\delta\tau = \frac{525}{\gamma}$ \cite{reason}. Thus, for a given $\beta$ we can obtain $\langle t\rangle$ from Eq. (\ref{Eq. 15}) and  thereby can obtain $\delta\tau$ by using Eq. (\ref{Eq. 17}).   
 
\begin{figure}[!ht]
\includegraphics[scale=0.23]{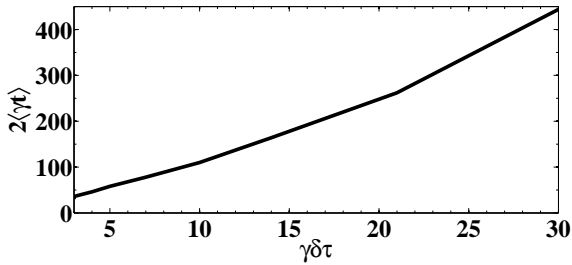}
\caption{The variation of 2$\langle \gamma t\rangle$ with the $\gamma\delta\tau$. Here, $\sigma = 2\pi \times 7.916 ~\mbox{kHz}$ and other parameters are same as in Fig. \ref{fig5}}
\label{fig7}
\end{figure}

We further  investigate the effect of changing the width of the Gaussian pulse on the above-mentioned correspondence. By increasing the width in frequency domain $(\sigma = 2\pi \times 7.916~\mbox{kHz})$, we found that the linearity between 2$\langle \gamma t\rangle$ and $\gamma\delta\tau$ is still maintained, [see Fig. \ref{fig7}].

We find that similar correspondence between $\langle t\rangle$ and $\delta\tau$ can be obtained  in alkali atoms in a tripod configuration \cite{dasgupta2002}, as well. Note that such configuration is known to work as a polarization splitter of pulses. 

Next, we discuss how this correspondence can be useful to obtain the number density ($N$) of the medium. The time $t_{-}$ ($t_{+}$) taken by $\sigma_{-}$ ($\sigma_{+}$) component of the pulse to reach the detector is related to DGD by $\delta\tau = t_{-} - t_{+} = L/v_{g}^{-} - L/v_{g}^{+}$, where, $v_{g}^{\pm}$ are the group velocities of $\sigma_{\pm}$ components inside the medium and are given by  Eq. (A. 12). Using the values of $\chi_{\pm}$ (Eqs. (A. 10), (A. 11)) , it can be easily shown that $\delta\tau \propto N$. Thus, we can get the value of $N$ by calculating $\delta\tau$ from the above mentioned correspondence. It is to be noted that $N$ could be calculated by using usual absorption method, in which the intensity of the input pulse gets reduced by a fraction $\xi = e^{-\alpha N}$ at the output. Here $\alpha = \frac{3L\lambda_{ij}^{2}}{2\pi}\mbox{Im}\left[\rho_{ij}^{(1)}\right]$. However, the associated relative error can be written as
\begin{equation}
\left|\frac{dN}{N}\right| = \left|\frac{1}{N\alpha}\right|\left|\frac{d\xi}{\xi}\right|
\label{Eq. 18}
\end{equation}
This implies that with decrease in the number density, the relative error in the number density increases for usual absorption procedure. In such a situation, when $N$ is small (i.e., when the absorption of the pulse is negligible), the proposed method of measuring $N$ using weak measurement would be much more useful.

\subsection{Effect of Absorption}
The above results are discussed for negligible absorption of the polarization components of the input field. But, in the presence of substantial absorption, the weak value (Eq. (\ref{Eq. 14})) can be written as
\begin{widetext}
\begin{equation}
\mbox{Re}W = \frac{\left(1 + \cot\left(\beta\right)\right)^{2} - \eta^{2}\left(1 - \cot\left(\beta\right)\right)^{2}}{\left(1 + \eta\right)^{2}\left[\cot^{2}\left(\beta\right)\sin^{2}\left(\frac{\omega_{0}\delta\tau}{2}\right) + \cos^{2}\left(\frac{\omega_{0}\delta\tau}{2}\right)\right]+\left(1 - \eta\right)^{2}\left[\cot^{2}\left(\beta\right)\cos^{2}\left(\frac{\omega_{0}\delta\tau}{2}\right) + \sin^{2}\left(\frac{\omega_{0}\delta\tau}{2}\right)\right] + 2\left(1 - \eta^{2}\right)\cot\left(\beta\right)}
\label{Eq. 19}
\end{equation}
\end{widetext}
where, $\eta$ gives the ratio of the electric field amplitude of $\sigma_{-}$ component relative to $\sigma_{+}$ component.  For equal absorption  $\left(\eta = 1\right)$, Eq. (\ref{Eq. 19}) reduces to Eq. (\ref{Eq. 16}).   In the case of significant absorption of $\sigma_{-}$ component $\eta \ll 1$ ,  Eq. (\ref{Eq. 19}) would give $\mbox{Re}W = 1$. On the contrary, when $\eta \gg 1$, i.e., when the $\sigma_{+}$ polarized component gets absorbed, the Eq. (\ref{Eq. 19}) would give $\mbox{Re}W = -1$. In these situations, the medium would act as a polarizing medium. This puts a limitation on the weak measurement of DGD in atomic vapor system in the presence of large absorption.  
\vspace{-0.5cm}
\section{Concluding Remarks}
The measurement of temporal separation can be limited by  several technical errors (e.g., misalignment of the polarizers, stability of laser pointing, the intensity resolution of the detector). Several schemes have been proposed to suppress these technical errors  in Refs. \cite{brunner2010,bruder2013,feizpour2011,starling2009,nishizawa2012,kedem2012}.   For our scheme to work, the conditions to be obeyed are $\langle t\rangle > \Delta t$, $\delta\tau \approx \Delta t$, where $\Delta t$ is the temporal resolution of the detector. For a detector with $\Delta t = 1~\mu$s, our proposal of measuring DGD would work well even with an ordinary detector. However, for a detector with higher temporal resolution \cite{sutter1999}, the condition $\delta\tau \approx \Delta t$ demands that $\delta\tau$ is to be reduced. Further, measurement of the temporal separation can be limited by statistical errors i.e. the total number of photons detected at the detector. The probability of successful post-selection is given by $p = |\langle\psi_{1}|\psi\rangle|^{2}$. In the regime where the two polarizers  are nearly orthogonal ($\beta\approx 0$), the intensity of the output pulse will be lowered ($I_{out} = \mbox{0.298}~ \mbox{nW}/\mbox{cm}^{2}$).   So, one needs to optimize for an appropriate value of $\beta$ by considering the detector efficiency. However, the overall sensitivity can be increased by using standard signal modulation and lock-in detection techniques \cite{hoston2008}.

In conclusion, we have discussed the use of weak values of polarization in temporally resolving two overlapping circular polarization components of a weak linearly polarized pulse, propagating through a coherently driven alkali atomic vapor. We find that while a post-selection would give rise to large weak value, the strong control field also plays a crucial role in obtaining a good temporal resolution. Such a control is inherently absent in experiments with photons passing through an optical fiber. We have shown that the temporal separation between the two polarization components can be obtained from the mean time of arrival of an output pulse in a postselected polarization. For negligible absorption, we obtain a linear relationship between this separation and the mean time, while in case of absorption, they exhibit a nontrivial relation as a function of control field Rabi frequency and the post-selected basis. We have provided full analytical and numerical results to demonstrate the above idea. 
\begin{acknowledgments}
One of us (S.D.) gratefully acknowledges Prof. G. S. Agarwal, who originally proposed the idea.
\end{acknowledgments}
\appendix*
\section{}
\begin{figure}[!ht]
\includegraphics[scale=0.2]{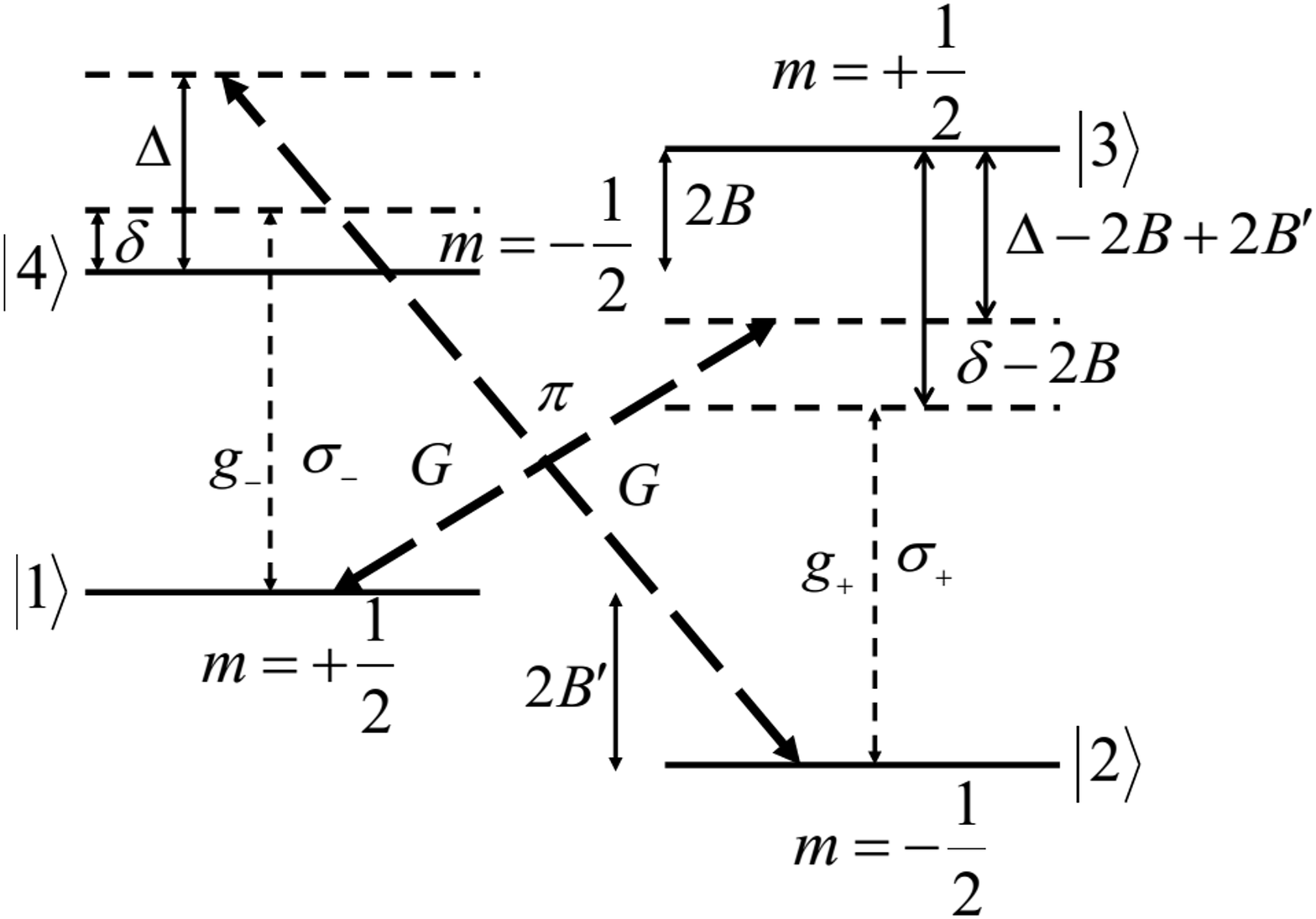}
\caption{Schematic energy-level structure of a four-level atomic system involving $J = \frac{1}{2} \leftrightarrow J = \frac{1}{2}$ transition. The level $|4\rangle$ ($|3\rangle$) is coupled to $|1\rangle$ ($|2\rangle$) by $\sigma_{-}$ ($\sigma_{+}$) component of the probe pulse with Rabi frequency $g_{-}$ ($g_{+}$). A $\pi$-polarized control field couple the level $|4\rangle$ ($|3\rangle$) to $|2\rangle$ ($|1\rangle$). The detuning of the $\sigma_{-}$ component from the respective transition is $\delta$ and the pumping field detuning is $\Delta$. The degeneracy of the excited states $|4\rangle$, $|3\rangle$ and the ground sub-levels $|1\rangle$, $|2\rangle$ have been removed by applying a dc magnetic field, so as to make the system an-isotropic. The corresponding Zeeman separations are $2B$ and $2B^{\prime}$, respectively.}
\label{fig10}
\end{figure}
We choose a generic four level configuration \cite{kumar2014} as shown in Fig. \ref{fig10}. A dc magnetic field is applied to remove the degeneracy of the excited and the ground states. In general, the Zeeman separation $2B$ of the excited states ($|3\rangle, |4\rangle$) is not the same as the Zeeman separation $2B^{\prime}$ of the ground manifolds ($|1\rangle, |2\rangle$), due to the difference in Land\'{e} g- factors in these manifolds. For $^{39}$K, $B^{\prime} = 3B$, $g_{e} = 2/3$ and $g_{g} = 2$ are the Land\'{e} $g$-factors of the excited and the ground sublevels. We consider propagation of a $\hat{x}$-polarized probe pulse (Eq. (\ref{Eq. 3})) through the atomic medium. The $\sigma_{\pm}$ components of this probe pulse interact with the transitions $|2\rangle \leftrightarrow |3\rangle$ and $|1\rangle \leftrightarrow |4\rangle$, respectively. The corresponding Rabi frequencies are given by $2g_{+} = 2\left(\frac{\vec{d}_{32}\cdot \hat{x}\varepsilon_{p}}{\hbar}\right)$ and  $2g_{-} = 2\left(\frac{\vec{d}_{41}\cdot \hat{x}\varepsilon_{p}}{\hbar}\right)$, where $\vec{d}_{ij}$ is the electric dipole moment matrix element between the levels $|i\rangle$ and $|j\rangle$. A strong $\pi$-polarized control field is applied to drive the transitions $|1\rangle \leftrightarrow |3\rangle$ and $|2\rangle \leftrightarrow |4\rangle$. The Rabi frequency of this control field is given by  $2G = 2\left(\frac{\vec{d}_{31}\cdot \hat{z}\varepsilon_{p}}{\hbar}\right) = 2\left(\frac{\vec{d}_{42}\cdot \hat{z}\varepsilon_{c}}{\hbar}\right)$.\\

The Hamiltonian for the above configuration can be written in the dipole approximation as
\begin{eqnarray}
\begin{array}{ccc}
\hat{H} &=&\hbar \left[ \omega_{21}\vert2\rangle\langle2\vert+\omega_{31}\vert3\rangle\langle3\vert+\omega_{41}\vert4\rangle\langle4\vert \right] \\
& & -\left[(\vec{d_{41}}\vert4\rangle\langle1\vert+\vec{d_{32}}\vert3\rangle\langle2\vert+h.c.).\vec{E_p}\right]\\
& & -\left[(\vec{d_{31}}\vert3\rangle\langle1\vert+\vec{d_{42}}\vert4\rangle\langle2\vert+h.c.).\vec{E_c}\right]\;.
\end{array}
\label{eq1}
\end{eqnarray}

Here zero of energy is defined at the level $\vert1\rangle$ and $\hbar\omega_{\alpha\beta}$ is the energy difference between the levels $\vert \alpha\rangle$ and $\vert \beta\rangle$.
Including the natural decay terms into our analysis and using the Markovian master equation under rotating wave approximation, the following density matrix equations are obtained:
\begin{widetext}
\begin{eqnarray}
\begin{array}{lll}
\dot{\tilde{\rho}}_{11} &=& \gamma_{13}\tilde{\rho}_{33}+\gamma_{14}\tilde{\rho}_{44}+i(G^\ast\tilde{\rho}_{31}-G\tilde{\rho}_{13})+i(g^{\ast}_{+}\tilde{\rho}_{41}e^{i\omega_{pc}t}-g_{+}\tilde{\rho}_{14}e^{-i\omega_{pc}t})\\
\dot{\tilde{\rho}}_{33} &=& -(\gamma_{13}+\gamma_{23})\tilde{\rho}_{33}+i(G\tilde{\rho}_{13}-G^\ast\tilde{\rho}_{31})+i(g_{-}\tilde{\rho}_{23}e^{-i\omega_{pc}t}-g^\ast_{-}\tilde{\rho}_{32}e^{i\omega_{pc}t})\\
\dot{\tilde{\rho}}_{44} &=& -(\gamma_{14}+\gamma_{24})\tilde{\rho}_{44}+i(G\tilde{\rho}_{24}-G^\ast\tilde{\rho}_{42})+i(g_{+}\tilde{\rho}_{14}e^{-i\omega_{pc}t}-g^\ast_{+}\tilde{\rho}_{41}e^{i\omega_{pc}t})\\
\dot{\tilde{\rho}}_{31} &=& i(\Delta - 2B + 2B^{\prime}+ +i\Gamma_{31})\tilde{\rho}_{31}+i(\tilde{\rho}_{11}-\tilde{\rho}_{33})G + i(g_{-}\tilde{\rho}_{21}-g_{+}\tilde{\rho}_{34})e^{-i\omega_{pc}t}\\
\dot{\tilde{\rho}}_{32} &=& i(\Delta - 2B +i \Gamma_{32})\tilde{\rho}_{32}+i(\tilde{\rho}_{12}-\tilde{\rho}_{34})G+i(1-\tilde{\rho}_{11}-2\tilde{\rho}_{33}-\tilde{\rho}_{44})g_{-}e^{-i\omega_{pc}t}\\
\dot{\tilde{\rho}}_{43} &=& (2iB - \Gamma_{43})\tilde{\rho}_{43}+i(G\tilde{\rho}_{23}-G^\ast \tilde{\rho}_{41})+i(g_{+}\tilde{\rho}_{13}e^{-i\omega_{pc}t}-g^\ast_{-} \tilde{\rho}_{42}e^{i\omega_{pc}t})\\
\dot{\tilde{\rho}}_{42} &=& i(\Delta +i \Gamma_{42})\tilde{\rho}_{42}+i(1-\tilde{\rho}_{11}-\tilde{\rho}_{33}-2\tilde{\rho}_{44})G +i(g_{+}\tilde{\rho}_{12}-g_{-}\tilde{\rho}_{43})e^{-i\omega_{pc}t}\\
\dot{\tilde{\rho}}_{41} &=& i( \Delta + 2B^{\prime} +i\Gamma_{41})\tilde{\rho}_{41}+i(\tilde{\rho}_{21}- \tilde{\rho}_{43})G+i(\tilde{\rho}_{11}- \tilde{\rho}_{44})g_{+}e^{-i\omega_{pc}t}\\
\dot{\tilde{\rho}}_{21} &=& (2iB^{\prime} - \Gamma_{21})\tilde{\rho}_{21}+i(G^\ast \tilde{\rho}_{41}-G \tilde{\rho}_{23})+i(g_{-}^\ast\tilde{\rho}_{31}e^{i\omega_{pc}t}-g_{+}\tilde{\rho}_{24}e^{-i\omega_{pc}t})\;,
\end{array}
\label{eq19}
\end{eqnarray}
\end{widetext}
where, $\Delta = \omega_{c} - \omega_{42}$ is the control field detuning from the transition $|2\rangle \leftrightarrow |4\rangle$, $\delta = \omega_{p} - \omega_{41}$ is the probe detuning from the transition $|1\rangle \leftrightarrow |4\rangle$, and $\omega_{pc} = \delta - \Delta - 2B^{\prime}$ is the probe-pump detuning.  $\gamma_{ij}$ is the spontaneous emission rate from the level $|j\rangle$ to $|i\rangle$, $\Gamma_{ij} = \frac{1}{2}\sum_{k} (\gamma_{ki} + \gamma_{kj}) + \gamma_{coll}$ is the dephasing rate of the coherence between the levels $|j\rangle$ and $|i\rangle$ and $\gamma_{coll}$ is the collisional decay rate. The transformations for the density matrix elements are as follows: $\rho_{31} = \tilde{\rho}_{31}e^{-i\omega_{c}t}$,    $\rho_{32} = \tilde{\rho}_{32}e^{-i\omega_{c}t}$, $\rho_{42} = \tilde{\rho}_{42}e^{-i\omega_{c}t}$,  $\rho_{41} = \tilde{\rho}_{41}e^{-i\omega_{c}t}$. The rest of the elements remain the same.\\
The steady state solutions of Eq. (\ref{eq19}) can be found by expanding the density matrix elements in terms of the harmonics of $\omega_{pc}$ as
\begin{equation}
\begin{array}{c}
\tilde{\rho}_{\alpha\beta}=\tilde{\rho}_{\alpha\beta}^{(0)}+g_{+}e^{-i\omega_{pc}t}\tilde{\rho}_{\alpha\beta}^{\prime(-1)}+g_{+}^\ast e^{i\omega_{pc}t}\tilde{\rho}_{\alpha\beta}^{\prime\prime(-1)}\\ +g_{-}e^{-i\omega_{pc}t}\tilde{\rho}_{\alpha\beta}^{\prime(+1)}+g_{-}^\ast e^{i\omega_{pc}t}\tilde{\rho}_{\alpha\beta}^{\prime\prime(+1)} \;.
\end{array}
\label{eq20}
\end{equation}

Thus, we obtain a set of algebraic equations of $\tilde{\rho}_{\alpha\beta}^{(n)}$. These equations can be solved for different values of $n$ to obtain following zeroth order solutions:
\begin{equation}
\begin{array}{ccc}
\tilde{\rho}_{33}^{(0)} = \frac{xy|G|^{2}\gamma_{14}}{Q},\\
\tilde{\rho}_{44}^{(0)} = \frac{xy|G|^{2}\gamma_{23}}{Q},\\
\tilde{\rho}_{11}^{(0)} = \frac{x}{Q}\gamma_{14}(\gamma_{13} + \gamma_{23} + y|G|^{2}),\\
\tilde{\rho}_{22}^{(0)} = \frac{y}{Q}\gamma_{23}(\gamma_{14} + \gamma_{24} + x|G|^{2}),\\
\tilde{\rho}_{13}^{(0)} = \frac{iG^{\ast}}{d_{1}^{\ast}}\frac{x\gamma_{14}}{Q}(\gamma_{13} + \gamma_{23}),\\
\tilde{\rho}_{24}^{(0)} = \frac{iG^{\ast}}{d_{2}^{\ast}}\frac{y\gamma_{23}}{Q}(\gamma_{14} + \gamma_{24}),
\end{array}
\end{equation}
where,
\begin{equation}
\begin{array}{c}
d_{1} = i(\Delta - 2B + 2B^{\prime}) - \Gamma_{31}; d_{2} = i\Delta - \Gamma_{42}; x = \frac{2\Gamma_{42}}{|d_{2}|^{2}}\\ y = \frac{2\Gamma_{31}}{|d_{1}|^{2}}; Q = \gamma_{23}(\gamma_{14} + \gamma_{24})y + \gamma_{14}(\gamma_{13} + \gamma_{23})x\\ + 2xy|G|^{2}(\gamma_{14} + \gamma_{23}).\\
\end{array}
\end{equation}
Thus, the probe coherence terms for $\sigma_{\pm}$ components can be written as
\begin{eqnarray}
\tilde{\rho}_{41}^{\prime(-1)} &=& A_{1} + B_{1} + C_{1},\\
\tilde{\rho}_{32}^{\prime(+1)} &=& A_{2} + B_{2} + C_{2}.
\end{eqnarray}
where,
\begin{equation}
\begin{array}{ccc}
A_{1} &=& \frac{-i\left[qrs + (q + r)|G|^{2}\right]\left(\tilde{\rho}_{11}^{(0)} - \tilde{\rho}_{44}^{(0)}\right)}{\left[pqrs + |G|^{2}(p + s)(q + r)\right] },\\
B_{1} &=& \frac{rsG\tilde{\rho}_{24}^{(0)}}{\left[pqrs + |G|^{2}(p + s)(q + r)\right]},\\
C_{1} &=& \frac{qsG\tilde{\rho}_{13}^{(0)}}{\left[pqrs + |G|^{2}(p + s)(q + r)\right]},\\
A_{2} &=& \frac{-i\left[uvw + (v + u)|G|^{2} \right]\left(\tilde{\rho}_{22}^{(0)} - \tilde{\rho}_{33}^{(0)}\right)}{\left[fuvw + |G|^{2}(f + w)(u + v)\right]},\\
B_{2} &=& \frac{vwG\tilde{\rho}_{13}^{(0)}}{\left[fuvw + |G|^{2}(f + w)(u + v)\right]},\\
C_{2} &=& \frac{uwG\tilde{\rho}_{24}^{(0)}}{\left[fuvw + |G|^{2}(f + w)(u + v)\right]}.

\end{array}
\end{equation}
with
\begin{equation}
\begin{array}{cc}

p = i\left(\omega_{pc} + \Delta + 2B^{\prime}\right) - \Gamma_{41}; ~ q = i\left(\omega_{pc} + 2B^{\prime}\right) - \Gamma_{21},\\
r = i\left(\omega_{pc} + 2B\right) - \Gamma_{43};~ s = i\left(\omega_{pc} - \Delta +2B\right) - \Gamma_{32},\\
f = i\left(\omega_{pc} + \Delta - 2B\right) - \Gamma_{32};~ u = i\left(\omega_{pc} - 2B^{\prime}\right) - \Gamma_{21},\\
v = i\left(\omega_{pc} - 2B\right) -\Gamma_{43};~ w = i\left(\omega_{pc} - \Delta - 2B^{\prime}\right) - \Gamma_{41}.
\end{array}
\end{equation}

The susceptibilities of the medium for the $\sigma_\pm$ components can be written as 
\begin{eqnarray}
\chi_{+} &=& \left(\frac{N|d_{32}|^2}{\hbar\gamma} \right)\tilde{\rho}_{32}^{\prime(+1)}\\
\chi_{-} &=& \left(\frac{N|d_{41}|^2}{\hbar\gamma}\right)\tilde{\rho}_{41}^{\prime(-1)}
\end{eqnarray}
The circular polarized components of the pulse travel with different group velocities inside the medium which are given by
\begin{eqnarray}
\begin{array}{ccc}
&v_{g}^{\pm} = \frac{c}{n_{g}^{\pm}}&\\
&=& \frac{c}{\left[1 + 2\pi \mbox{Re}\left[\chi_{\pm}(\omega)\right] + 2\pi\omega \frac{\partial }{\partial \omega}\mbox{Re}\left[\chi_{\pm}(\omega)\right]\right]}_{\omega = \omega_{0}}
\end{array}
\end{eqnarray}

\bibliographystyle{elsarticl-num}

\begin{thebibliography}{50}
\bibitem{reason} The disparity between the measured and the calculated values of $\gamma\delta\tau$ can be attributed to the non-Gaussian pulse shape at the output in Fig. \ref{fig6}.
\end{thebibliography}

\end{document}